\documentclass[12pt,preprint]{aastex} 
  
\newcommand{\cm  }{\,{\rm cm}^{-3} } 
\newcommand{\etal }{{et al.} }
\newcommand{\msun}{\thinspace M_\odot}

\shorttitle{} 
\shortauthors{}

\newcommand\lsim{\mathrel{\rlap{\lower4pt\hbox{\hskip1pt$\sim$}}
        \raise1pt\hbox{$<$}}}
\newcommand\gsim{\mathrel{\rlap{\lower4pt\hbox{\hskip1pt$\sim$}}
        \raise1pt\hbox{$>$}}}
 
\shorttitle{Star Formation in a Relic HII Region}
\shortauthors{Machida  \etal 2009}

\begin{document} 
  
\title{Star Formation in Relic HII Regions of the First Stars: 
Binarity and Outflow Driving} 
\author{
Masahiro N. Machida\altaffilmark{1,2},
Kazuyuki Omukai\altaffilmark{1}, and 
Tomoaki Matsumoto\altaffilmark{3}} 
\altaffiltext{1}{National Astronomical Observatory of Japan, 
Mitaka, Tokyo 181-8588, Japan; omukai@th.nao.ac.jp} 
\altaffiltext{2}{Department of Physics, Kyoto University, 
Kyoto 606-8502, Japan; machiam@scphys.kyoto-u.ac.jp} 
\altaffiltext{3}{Faculty of Humanity and Environment, Hosei University, 
Chiyoda, Tokyo 102-8160} 
 
\begin{abstract} 
Star formation in relic HII regions of the first stars is investigated using magneto-hydrodynamical simulations with a nested grid method that
covers $\sim10$ orders of magnitude in spatial scale and $\sim 20$ orders of magnitude in density contrast.
Due to larger fraction of H$_2$ and HD molecules, its prestellar thermal evolution is considerably different from that in the first star formation.
Reflecting the difference, two hydrostatic cores appear in a nested manner: 
a protostar is enclosed by a transient hydrostatic core, which appears during 
the prestellar collapse.
If the initial natal core rotates fast at a rate with rotational to 
gravitational energy ratio $\beta_0\ga 0.01-0.1$, 
the transient hydrostatic core fragments to $\sim10\msun$ sub-cores
at density $\sim10^9\cm$.
With smaller rotation energy, fragmentation occurs at higher density
while a single protostar forms without fragmentation if rotation is 
extremely slow with $\beta_0 \la 10^{-6}-10^{-5}$.
If magnetic field is present, these threshold values of $\beta_0$ is 
boosted owing to angular momentum transport by the magnetic breaking.
Magnetic field also drives the protostellar outflows.
With strong magnetic field, two distinct outflows are observed:
The slower one emanates from the transient hydrostatic core, 
while the faster one from the protostar. 
These flows may affect the final stellar mass
by ejecting some of masses in the initial core, and
also may play some role in driving and maintenance of 
interstellar turbulence in young galaxies.
\end{abstract}  
 
\section{Introduction}
Theoretical studies in the last decade revealed that
first stars in the universe were born in redshift $z\sim 20-30$ as
very massive stars with $\ga 100M_{\sun}$ (e.g., Ciardi \& Ferrara 2005).
It is also suggested that a fraction of them are formed as binaries 
or multiples as long as their natal cores somewhat rotate 
(Machida et al. 2008a; Turk, Abel, \& O'Shea 2009). 
To drive protostellar outflows, ubiquitously observed 
in the solar neighborhood and 
affecting the protostellar accretion, 
from the first stars, however, requires 
a magnetic field of $10^{-8}-10^{-9}$G at $10^{3}{\rm cm^{-3}}$ 
(Machida et al. 2006a), 
higher than the level generated by
mechanisms currently proposed, e.g., Biermann batteries at 
the cosmic recombination or halo virialization 
(Ichiki et al.; Xu et al. 2008) by 2-3 orders of magnitudes.
Hence the outflows/jets are hard to emerge at least 
in the early phase of protostellar accretion. 

Even in the primordial gas, prestellar thermal evolution, 
which affects fragmentation and thus characteristic stellar mass, 
can differ in a cloud disturbed by a former generation of stars. 
The stars born in such an environment begins to be 
recognized as a different class of objects and 
called Population III.2 stars, 
while the genuine first stars are Population III.1 stars (O'Shea et al. 2008).
Various feedbacks can affects cradles of subsequent 
generation of stars, and thus Pop III.2 in principle includes 
distinct sub-classes.
The most studied example, though, is those formed in relic HII 
regions of the first stars.
 
A large HII region forms around a massive first star.
Pop III stars with mass $\gtrsim 100M_{\sun}$ end their lives 
as supernovae only in a limited mass range ($140-260M_{\sun}$; 
Heger \& Woosley 2002; Umeda \& Nomoto 2002), 
and thus the most of them would collapse directly as black holes
rather quiescently.
The surrounding gas is left ionized and starts recombining 
gradually, and another episode of star formation commences
after $\sim 100$ Myrs.
In such relic HII regions, abundant electron, which is the catalyst for 
H$_2$ formation, allows the rapid replenishment of H$_2$ to
the fraction $\sim 10^{-3}$, about an order of magnitude higher than 
in the Pop III.1 case.
The cooler environment ($\la 150$K) due to the higher H$_2$ cooling 
is preferable for HD formation, and its cooling further lowers 
the temperature down to a few tens of Kelvins, close to the CMB 
temperature (Nagakura \& Omukai 2005).
The fragmentation mass-scale due to the HD cooling is about 
$\sim 40M_{\sun}$, and most of it 
is expected to be accreted by the formed star eventually
(Yoshida, Omukai, \& Hernquist 2007; McGreer \& Bryan 2008).
Not only in the relic HII regions, but the HD cooling can be 
important also in a medium ionized at the virialization of 
a large halo or even in a small halo if dynamical heating 
during mass assembly is weak (Ripamonti 2007).
This ``HD mode'' of primordial star formation 
may be a more common mode than the Pop III.1 one 
in forming galaxies.

In addition, if all the Pop III.1 stars were 
indeed very massive with a few hundreds of solar mass, 
they might have been very inefficient cauldrons for metals 
owing to the limited mass range for supernovae.
With smaller mass and being able to explode as Type II supernovae, 
Pop III.2 stars could be more responsible for the metals 
in the early universe.
To predict its mass range more accurately, 
we need to know the conversion efficiency from 
a dense core to stars, which is determined by 
binarity/multiplicity, feedback from protostellar outflow etc.
As an approach to this problem, 
here, we study the prestellar collapse of rotating dense cores
that is produced in an initially ionized gas, putting main 
emphasis on the condition of binary formation and launching 
of outflows.

This paper is organized as follows.
The numerical method is briefly described in \S 2, 
and the results are presented in \S3.
Finally, we discuss its significance in the cosmological context 
and caveats in \S 4. 

\section{Numerical Method}
First we calculate the temperature evolution by using a free-falling 
one-zone model of Omukai et al. (2005), starting from  
number density $1{\rm cm^{-3}}$, temperature 7000K 
and the initial ionization degree 0.1.
The CMB radiation at 27.3K is imposed.
With given temperature evolution as a function of density, 
we performed magneto-hydrodynamics (MHD) by a 
nested-grid method to gain high resolution near the center.
As the initial condition, we take the critical Bonner-Ebert sphere 
with central density $10^{4}{\rm cm^{-3}}$ but with a factor 1.68 
of density enhancement to induce the gravitational instability.
In addition, $m=2$ mode of density perturbations are
imposed with $10\%$ to break axi-symmetry and cause fragmentation.  
The initial angular momentum and magnetic field are 
parallel to the z-axis, and are parameterized by
the ratio of rotation to gravitational energy, $\beta_{0}$,  
and the ratio of magnetic to gravitational energy $\gamma_{0}$, 
respectively.
These quantities are related to the initial rotation rate of the core 
$\Omega_{0}\,(\rm s^{-1})$ and magnetic flux density $B_{0}\,(\rm G)$ 
as 
$\beta_{0}=0.1 [\Omega_{0}/3 \times 10^{-14}(\rm s^{-1})]^{2}$
and
$\gamma_{0}=1.5 \times 10^{-5} [B_{0}/10^{-7}(\rm G)]^{2}$, respectively.
The magnetic field is assumed to be frozen to the gas.
For details of the MHD and initial conditions, see Machida et al. (2005). 
The initial parameters of calculated runs and 
summary of the result are presented in Table 1.  

\section{Results}
The adopted temperature evolution in the HD mode is shown in Figure 1. 
Also shown is that in the Pop III.1 case (``H$_2$ mode''), where the 
collapse begins from an almost neural condition.
The temperature in the HD mode is lower than that 
in the H$_2$ mode for $10^{2-7}{\rm cm^{-3}}$, 
and reaches several 10K owing to the HD cooling.
The HD-cooling rate saturates at $\sim 10^{5}{\rm cm^{-3}}$ by 
collisional de-excitation, and the temperature begins to increase 
with density.
The fragmentation scale $\sim 40M_{\sun}$ is set around there 
(Yoshida et al. 2007). 
At $10^{8}{\rm cm^{-3}}$, the H$_2$ begins to form via the three-body 
reaction, whose heating makes the temperature jump up abruptly. 
The simultaneous increase of pressure slows down the gravitational 
collapse near the center of the dense core, and transient hydrostatic 
core forms just for a brief period. 
The core begins to collapse dynamically again until the permanent hydrostatic
core, i.e., the protostar, forms at $\ga 10^{21} {\rm cm^{-3}}$ 
unless the centrifugal supports prevents the collapse before that.
Similar transient hydrostatic objects, the so-called first cores, 
form in metal-enriched cores when their central parts become optically 
thick to the dust absorption.

In metal-enriched cases, the first-core formation is an important epoch 
for fragmentation of rotating cores (Machida et al. 2009).
In the isothermal phase, i.e., before the first-core formation, 
even though the core becomes flattened by rotation, its collapse
does not significantly slow down until the adiabatic temperature-increase 
leading to the first-core formation.
Shortly after the first-core formation, if the 
core is flat enough, the slow contraction allows perturbations 
to grow, thereby causing fragmentation into a binary/multiple system. 
On the other hand, in the H$_2$ mode of primordial star formation 
(i.e., PopIII.1 formation), where the temperature continues to increase 
gradually without transient-core formation, no characteristic epoch 
for binary/multiple formation exists.
In this case, when a thin disk forms by rotation, 
it fragments after further contraction of a few orders 
of magnitude in density. 

The final states of the calculations are presented for 
different combinations of rotation ($\beta_{0}$) and 
magnetic field strength ($\gamma_{0}$) in Figure 2. 
The cores with initially large rotation 
($\beta_{0} \ga 0.01-0.1$ in the $\gamma_{0}=0$ cases) fragment shortly 
after the transient core formation and form wide proto-binaries with 
initial masses $\sim 1-10 M_{\sun}$ and 
separations $\sim 10^{4}$ AU (cases {\it a} in Figure 2).    
Those with smaller rotation ($10^{-5} \la \beta_{0} \la 10^{-2}$ 
in the $\gamma_{0}=0$ cases) fragment at higher density 
before the protostar formation
unless the rotation is extremely small 
($\beta_{0} \la 10^{-6}-10^{-5}$ in the $\gamma_{0}=0$ cases).
Such high-density ($10^{17-20} {\rm cm^{-3}}$) fragmentations
produce close proto-binaries with initial masses $\sim 10^{-2} M_{\sun}$ 
and separations $\sim 10^{-2}$ AU (cases {\it b} in Figure 2).
Note that some models classified in case {\it b} show merger of fragments; 
i.e., models ($\beta_0$, $\gamma_0$) = ($10^{-2}$, $1.5\times10^{-5}$) and 
($10^{-4}$, $1.5\times10^{-5}$).
We expect that the merged cores finally evolve in a single massive star, 
since spiral structures in them effectively transfer 
the angular momenta outward, which prevents further fragmentation.
Those with even smaller rotation collapse all the way without 
fragmentation and form single protostars (cases {\it c} in Fig.~2).
While the collapse is approximately spherical,  
the rotation parameter $\beta$ at the center of the core increases as 
$\propto n^{1/3}$, where $n$ is the number density, from the initial value 
$\beta_{0}$.
When $\beta$ reaches 0.04-0.1, the radial contraction is retarded 
by the centrifugal force. The core becomes a thin disk 
after more contraction by a few orders of magnitude in density 
(e.g., Machida et al. 2008). 
The threshold $\beta_{0}$ values (i.e., $0.01-0.1$ and $10^{-6}-10^{-5}$) 
for binary formation correspond to the conditions of thin-disk formation 
before formation of the transient core (at $\sim 10^{8}{\rm cm^{-3}}$) and 
the protostar (at $\sim 10^{21}{\rm cm^{-3}}$), respectively.

The magnetic field reduces the angular momentum of the cores 
through magnetic breaking, hence boosting the threshold rotation rate 
for fragmentation.
For example, with initial magnetic field $\gamma_0=1.5\times 10^{-3}$ 
($1.5\times 10^{-1}$), the threshold $\beta_{0}$ 
for fragmentation increases above $10^{-4}$ ($10^{-2}$, respectively)
from $\sim 10^{-5}$ in the $\gamma_{0}=0$ case.

Although the initial mass of fragments is very small, 
they are embedded in a dense core and subsequently grow by accretion 
until the majority of the core material ($\la 10M_{\sun}$) is 
accreted by the binary/multiple system (Yoshida et al. 2008).
However, if launched, protostellar outflow may reduce 
the conversion efficiency from the core to the system 
by evacuating some of envelope materials. 

When the initially ordered magnetic field is highly twisted by rotation, 
outflows are launched either by magnetic centrifugal force 
or by pinching by magnetic pressure.
During dynamical collapse, the cores do not rotate so many times
as to twist the magnetic field enough.
After hydrostatic core formation, the magnetic field is amplified
and distorted owing to the slow contraction, 
thereby enabling the launching of slow outflows. 
In the case of present-day star formation, two kinds of outflows 
appear, corresponding to two epochs of hydrostatic core formation, 
i.e., first core and protostar (Machida et al. 2008b).
In our case, owing to the presence of the transient hydrostatic core, 
similar two flows are launched, i.e., 
the slower one (2-3km/s) at transient core formation, and 
the faster one (20-40km/s) just at the protostar formation. 
The speeds of the flows increase with the mass of driving objects, 
i.e., the transient hydrostatic core and protostar.
According to our experiments, 
the minimum initial magnetic field needed for driving 
the fast (slow) flow was about $10^{-8}$G ($10^{-7}$G, respectively). 
On the other hand, in the Pop III.1 case, 
only fast flow is launched
as no transient hydrostatic stage exists (Machida et al. 2006a).

\section{Discussion}
Currently, we have no information about rotation rates of 
Pop III.2 dense cores. 
If we assume that they are similar to the values in the Milky Way
($\beta_{0} \sim 10^{-4}-0.07$ with the median of 0.02; Caselli et al. 2002), 
fragmentations typically occur at far higher densities than the 
transient core formation (i.e., case {\it b} in Figure 2), 
and produce close binary/multiple systems.
Since the mass scale of the natal cores is $30-40M_{\sun}$, 
set by fragmentation induced by the HD cooling 
(Yoshida et al. 2007), 
the final product would be a close binary of $10-20M_{\sun}$ stars.

For the Pop III.1 protostars, only the primordial magnetic field of
the cosmological origin is available.
On the other hand, the Pop III.2 stars are able to use additional 
magnetic fields, which might have been generated and amplified 
by activities of previous generations of stars, 
including both the Pop III.1 and III.2.
Therefore, the Pop III.2 protostars could launch the outflow 
even though an order-of-magnitude larger magnetic field is required 
for the fast-flow launching than in the Pop III.1 case.
To know whether the protostellar jet is launched, 
we need to wait future numerical simulations to narrow down 
the possible values for magnetic field in such an environment.
If indeed launched, the jet expels some materials in the natal 
core, and hence reducing the final mass in the binary system.
Matzner \& McKee (2000) estimated the efficiency from the core 
to the system at $25-70\%$ after the jet feedback 
in the present-day star formation
by way of hydrodynamical simulation. 
If this efficiency is similar in our case, the final system mass 
would be $20M_{\sun}$ out of a $40M_{\sun}$ core.
To know the final mass of stars, 
we need to follow evolution in the accretion phase, which is 
left for future studies.

As mentioned in \S2, we used the temperature evolution as a function 
of density, which is calculated by the one-zone model 
of Omukai et al. (2005), where spherical clouds with 
the Jeans lengths are assumed to collapse at about the free-fall rate. 
In reality, thermal evolution depends on the cloud dynamics, and 
vice versa.
For example, when the cloud shape deviates from the sphere, 
the optical depths and thus the cooling rates can differ
from the assumed values.  
Also, slower collapse leads to lower compressional heating.
Moreover, our method cannot capture the shock heating properly.
This could be a significant drawback in simulating more turbulent 
situations (see Clark \etal 2008 for more discussions).

The first galaxy, harboring the first star clusters, 
is found to be turbulent owing to the cold accretion flow 
at its formation by numerical experiments (Wise \& Abel 2007; 
Greif et al. 2008). 
Here, however, we have not use turbulent initial conditions for simplicity.
If the turbulent energy is comparable to the gravitational one, 
the cloud evolution can be significantly affected in particular 
in early phases leading to formation of dense cores 
(e.g., Clark \etal 2008).
Rotational energy of the dense cores would be set by the turbulence and 
tend to be higher in the more turbulent environment.
In this study, we have regarded the degree of rotation as a free parameter 
and followed the evolution only after the formation of the 
dense cores.

The protostellar outflows by Pop III.2 stars could contribute 
to drive and maintain the turbulence in the first galaxies.
However, this phase might be short-lived as the first galaxies 
will be metal-enriched before long. 
In addition, such metal-free first galaxies might be rare as 
the previous generations of stars have already polluted with metals.
Therefore, protostellar outflows by the Pop II stars would be more 
important in this context and awaits future studies 
(e.g., Machida et al. 2009).

\acknowledgements 
Numerical computation in this work was carried out at the 
Yukawa Institute Computer Facility.
This study is supported in part by the Grants-in-Aid
by the Ministry of Education, Science and Culture of Japan 
(21740136:MM, 19047004, 21684007:KO, 20540238:TM).

\begin{table}
\caption{Models and Results
}
\label{table}
\begin{center}
\begin{tabular}{c|cccc|ccccccc} \hline
{\footnotesize Model} & $\beta_0$ & $\gamma_0$ & $\Omega_0$ {\scriptsize [s$^{-1}$]}& $B_0$ {\scriptsize [G]}  &  
 $n_{\rm f}$ {\scriptsize ($\cm$)}$^a$ & Outflow$^b$ 
\\ \hline
1  & 0 & 0.15                      & 0  & $10^{-5}$           & --- & None \\
2  & 0 & $1.5\times10^{-3}$         & 0  & $10^{-6}$           & --- & None \\
3  & 0 & $1.5\times10^{-5}$         & 0  & $10^{-7}$           & --- & None \\
\hline
4  & 0.1 & 0.15                    & $3.0\times10^{-14}$  & $10^{-5}$ & $1.3\times10^{18}$ & Both \\
5  & 0.1 & $1.5\times10^{-3}$       & $3.0\times10^{-14}$  & $10^{-6}$ & $6.7\times10^{11}$ & Slow \\
6  & 0.1 & $1.5\times10^{-5}$       & $3.0\times10^{-14}$  & $10^{-7}$ & $1.5\times10^9$ & None \\
7  & 0.1 & 0                       & $3.0\times10^{-14}$  & 0         & $1.5\times10^9$ & None \\
\hline
8 & $10^{-2}$ & 0.15              & $9.4\times10^{-15}$  & $10^{-5}$ & --- & Slow \\
9 & $10^{-2}$ & $1.5\times10^{-3}$ & $9.4\times10^{-15}$  & $10^{-6}$ & $5.4\times10^{17}$ & Both \\
10 & $10^{-2}$ & $1.5\times10^{-5}$ & $9.4\times10^{-15}$  & $10^{-7}$ & $2.4\times10^{17}$ & Fast \\
11 & $10^{-2}$ & 0                 & $9.4\times10^{-15}$  & 0         & $3.7\times10^{17}$ & None \\
\hline
12 & $10^{-3}$ & 0.15              & $3.0\times10^{-15}$  & $10^{-5}$ & --- & Fast \\
13 & $10^{-3}$ & $1.5\times10^{-3}$ & $3.0\times10^{-15}$  & $10^{-6}$ & $2.1\times10^{20}$ & Fast \\
14 & $10^{-3}$ & $1.5\times10^{-5}$ & $3.0\times10^{-15}$  & $10^{-7}$ & $1.7\times10^{20}$ & Fast \\
15 & $10^{-3}$ & 0                 & $3.0\times10^{-15}$  & 0         & $1.9\times10^{20}$ & None \\
\hline
16 & $10^{-4}$ & 0.15              & $9.4\times10^{-16}$  & $10^{-5}$ & --- & Fast \\
17 & $10^{-4}$ & $1.5\times10^{-3}$ & $9.4\times10^{-16}$  & $10^{-6}$ & $1.2\times10^{22}$ & Fast\\
18 & $10^{-4}$ & $1.5\times10^{-5}$ & $9.4\times10^{-16}$  & $10^{-7}$ & $2.5\times10^{20}$ & Fast\\
19 & $10^{-4}$ & 0                 & $9.4\times10^{-16}$  & 0         & $2.9\times10^{20}$ & None\\
\hline
20 & $10^{-5}$ & 0.15              & $3.0\times10^{-15}$  & $10^{-5}$ & --- & Fast\\
21 & $10^{-5}$ & $1.5\times10^{-3}$ & $3.0\times10^{-15}$  & $10^{-6}$ & --- & Fast\\
22 & $10^{-5}$ & $1.5\times10^{-5}$ & $3.0\times10^{-15}$  & $10^{-7}$ & $2.2\times10^{21}$ & Fast\\
23 & $10^{-5}$ & 0                 & $3.0\times10^{-15}$  & 0         & $2.4\times10^{21}$ & None\\
\hline
24 & $10^{-6}$ & 0.15              & $9.4\times10^{-17}$  & $10^{-5}$ & --- & None\\
25 & $10^{-6}$ & $1.5\times10^{-3}$ & $9.4\times10^{-17}$  & $10^{-6}$ & --- & None\\
26 & $10^{-6}$ & $1.5\times10^{-5}$ & $9.4\times10^{-17}$  & $10^{-7}$ & --- & None\\
27 & $10^{-6}$ & 0                 & $9.4\times10^{-17}$  & 0         & --- & None\\
\hline
\hline
\end{tabular}
\end{center}
$^a$ Number density at fragmentation epoch\\
$^a$ Emergence of Slow, Fast and Both (slow and fast) outflows. 
\end{table}

\newpage
\bigskip

\plotone{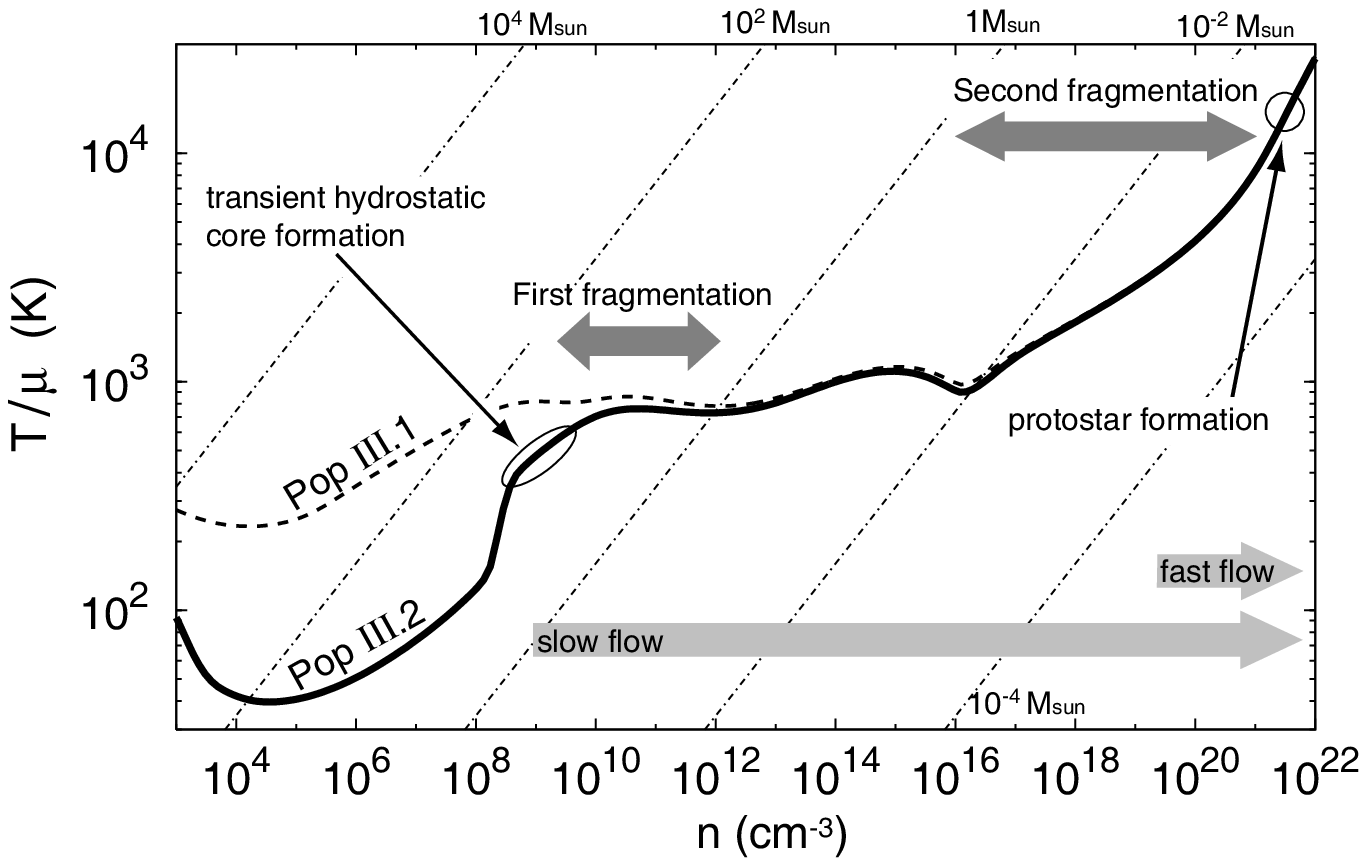} 
\figcaption[]{
Thermal evolution of a metal-free prestellar core from an initially 
ionized state as function of the number density 
(thick solid; indicated as ``Pop III.2'').
The temperature divided by mean molecular weight $\mu$ is shown.
Also shown for comparison is that for the Pop III.1 case 
(dashed).
The lines of constant Jeans mass are presented by the thin dot-dashed line. 
Some characteristic epochs are indicated by arrows with captions. 
}
\clearpage

\begin{figure}
\vspace{-0.5cm}
\includegraphics[width=140mm]{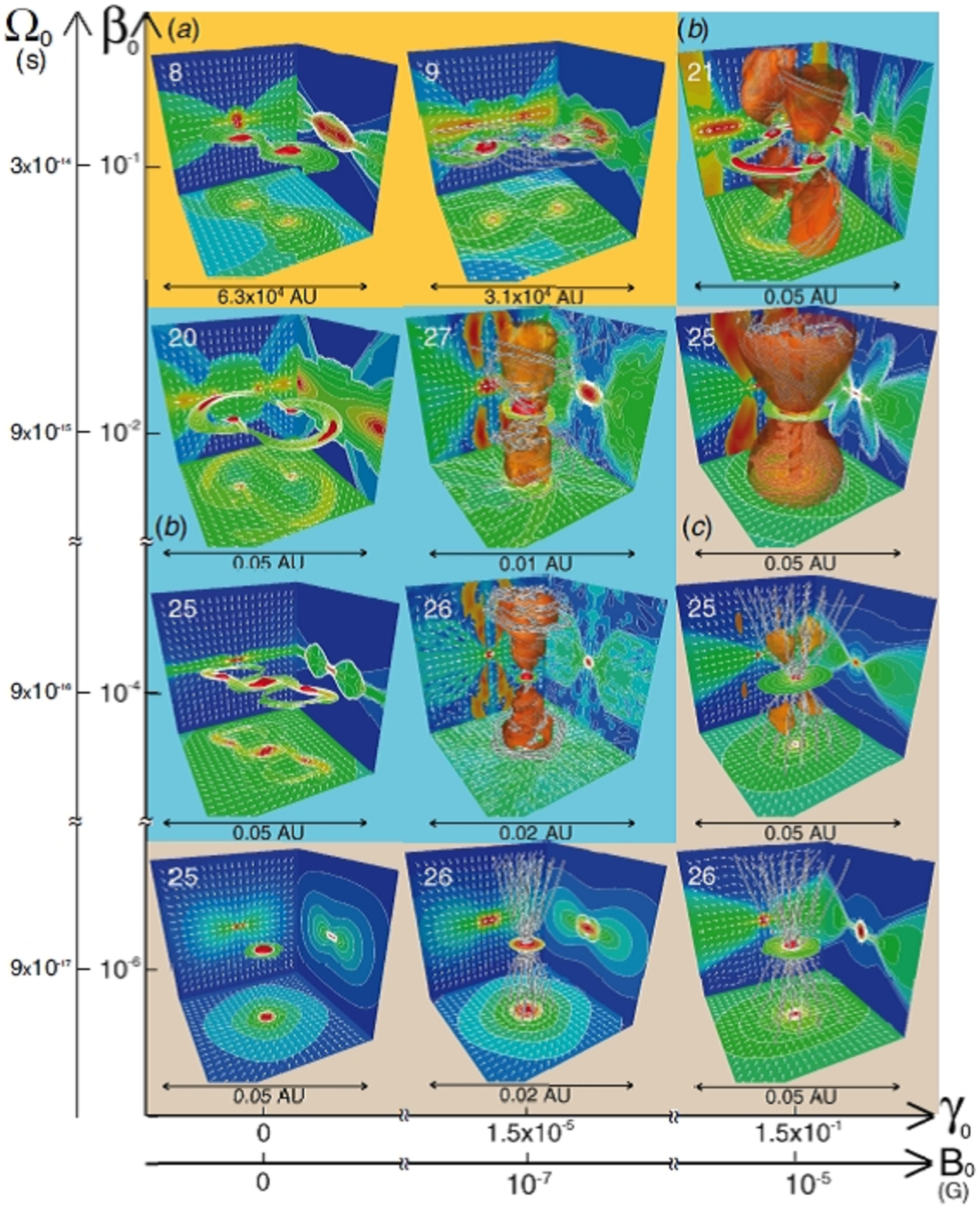}
\caption{
{\small
The final states for 
runs with different rotation ($\beta_{0}$) and magnetic field strength 
($\gamma_{0}$). The results are classified according to fragmentation 
properties; (a) fragmentation shortly after the transient core formation,
(b) fragmentation at higher density, and (c) no fragmentation, 
which are indicated by the background colors.
Note the larger binary separations ($\sim 10^{4}$ AU) 
in the cases ({\it a}) than those in ({\it b}; $\sim 10^{-2}$ AU).
In each panel, the structure of high-density region ({\it red isodensity surface}) and magnetic field lines 
({\it black and white streamlines}), and outflow shape 
(the region of radial velocity $v_r>0$; {\it transparent orange surface}) 
are plotted in three dimensions, while the density contour ({\it color and contour lines}) 
and velocity vectors ({\it arrows}) are projected on each wall surface.
The grid level $l$ and grid scale are shown at the upper-left corners and bottom, 
respectively, of each panel.
}}
\end{figure}
\end{document}